\documentclass[reprint,amsmath,amssymb,aps,prl]{revtex4-2}
\usepackage{graphicx}% Include figure files
\usepackage{dcolumn}% Align table columns on decimal point
\usepackage{bm}% bold math
\usepackage{mathtools}
\begin{document}
\preprint{APS/123-QED}
\title{Unexpected Robustness of Multipartite Entanglement against\\Sudden Death from Spontaneous Emission}
\author{Songbo Xie}
\email{sxie9@ur.rochester.edu}
\author{Daniel Younis}
\author{Joseph H. Eberly}
\affiliation{Center for Coherence and Quantum Optics, and Department of Physics and Astronomy, University of Rochester, Rochester, New York 14627, USA}

\date{\today}
\begin{abstract}
Entanglement is known to decay even in isolated systems, an effect attributed to spontaneous emission. This fragility of entanglement can be exacerbated by entanglement sudden death (ESD), where entanglement drops to zero abruptly within a finite time. It is natural to assume that multipartite entanglement is more vulnerable to ESD, as it involves more parties experiencing spontaneous emission. In this work, we challenge this assumption and present a contrasting conclusion, asserting that multipartite entanglement demonstrates increased robustness against ESD from spontaneous emission.

\end{abstract}
\maketitle
{\it Introduction}.---Entanglement is a fundamental concept in quantum information and computing. Its power can be enhanced when multiple parties are involved, leading to the emergence of a new criterion named ``genuine multipartite entanglement'' (GME). GME exists only when a system's state is nonbiseparable, indicating that it cannot be expressed as a tensor product of two distinct subsystems \cite{ma2011,xie2021}. GME is essential for various quantum tasks such as quantum error correction \cite{laflamme1996}, quantum metrology \cite{chin2012}, and quantum teleportation \cite{karlsson1998}. Thus, comprehending GME is crucial for advancing and enhancing quantum technologies. 

However, entanglement can be fragile, as it is susceptible to decay even in a vacuum, due to the system's spontaneous emission. Intriguingly, Yu and Eberly discovered that under certain circumstances, spontaneous emission can cause entanglement to vanish completely and also abruptly, {\it i.e.}, within a finite time, a phenomenon termed entanglement sudden death (ESD) \cite{yu2004,yu2009}. Considering the increased complexity of GME, which involves more parties experiencing spontaneous emission, it seems natural to assume that GME is at a greater risk of experiencing ``sudden death.''

In this study, we challenge this assumption and present a counter-intuitive finding that ESD can be less frequent in systems involving a greater number of parties. The fact that this result has remained undiscovered for many years since the first revelation of ESD is not entirely surprising. There are two primary reasons that account for this delayed realization.

The first reason is the absence of the ``genuine'' consideration in previous multipartite entanglement measures. For instance, the widely used ``three-tangle'' entanglement measure, proposed by Coffman, Kundu, and Wootters \cite{coffman2000}, fails to detect entanglement for the $W$ states, an important class of genuinely entangled states. Another example can be the generalized concurrence measure presented in \cite{carvalho2004}, which has also been employed to study multi-party ESD. Unfortunately, this measure fails to fullfil the ``genuine'' requirement, as it assigns entanglement to biseparable states. Utilizing an entanglement measure that is not genuine can lead to incorrect predictions of ESD or the absence of ESD.

The second reason for the delayed realization is that evaluating entanglement for mixed states in multi-party systems presents a formidable computational challenge. Existing valid investigations of ESD heavily rely on entanglement measures with closed forms for mixed states, with concurrence and negativity the only known examples. But these two measures are only applicable to small systems. Specifically, concurrence has a closed form exclusively for two-qubit systems \cite{hill1997}. On the other hand, the Peres–Horodecki criterion, on which negativity is based, has been shown to be sufficient only for $2\times2$ and $2\times3$ dimensional systems \cite{peres1996}. Consequently, for larger systems, the negativity measure derived from this criterion only provides a lower-bound estimation of the entanglement. This implies that a system can still be entangled when ESD is predicted using negativity. 

The aforementioned two reasons have hindered researchers from uncovering the true origin of ESD. In this work, we successfully overcome these obstacles and demonstrate genuine ESD for a three-qubit system subjected to spontaneous emissions. By identifying a specific ``trigger'' state, we subsequently establish that ESD may occur less frequently in systems with a greater number of qubits.

{\it Evaluation of GME}.---To account for the ``genuine'' element, we introduce a geometric measure of GME for three-qubit systems, referred to as concurrence fill \cite{xie2021,xie2022}. This measure associates a triangle with each three-qubit state and demonstrates that the triangle's area serves as a proper GME measure. Recently, the concurrence fill measure has been extended to quantify GME in larger systems as well \cite{xie2023}.

The concurrence fill measure was initially defined solely for pure states. In order to extend its applicability to mixed states, a convex-roof construction is required. This extension is sometimes called ``entanglement of creation'', to emphasize the amount of entanglement needed to create a given entangled state \cite{bennett1996}. The construction is expressed through an optimization formula
\begin{equation}\label{convexhull}
    \begin{split}
        F(\rho)=\inf_{\{p_i,\psi_i\}}\sum_ip_iF(\psi_i).
    \end{split}
\end{equation}
Here, the infimum is taken over all possible decompositions $\rho=\sum_ip_i|\psi_i\rangle\langle\psi_i|$, and $F(\psi_i)$ is the measure for pure states. A mixed state $\rho$ contains no genuine entanglement if and only if $F(\rho)=0$.

The complexity of Eq.~\eqref{convexhull} is evident, as the number of terms appearing in the sum cannot be predetermined. In principle, an infinite number of parameters are required to fully characterize this optimization. Although a new technique has recently been proposed for estimating this value experimentally \cite{xie2023estimate}, precise numerical evaluation of this formula remains a significant challenge and may even be considered impossible. It is worth noting that two different approaches have been proposed to evaluate this formula, but they have not gained widespread recognition.

The first approach was proposed by R\"othlisberger, Lehmann, and Loss \cite{rothlisberger2009}. They discovered that, for a rank-$r$ density matrix $\rho$, the result is already highly accurate when a cutoff is chosen such that the number of pure-state terms in the decomposition Eq.~\eqref{convexhull} does not exceed $r+4$. Consequently, the total number of parameters is given by $r(r+8)$. In our own experience, we have verified that the $r+4$ cutoff indeed yields sufficient accuracy for addressing the ESD problem, even when the system's entanglement is remarkably small---on the order of $10^{-3}$, as will be shown later. For the parametrization, we employ the Euler-Hurwitz angles, a technique introduced in Ref.~\cite{rothlisberger2009}. 

Nonetheless, this {\it cutoff} approach encounters limitations when the rank of the mixed state $\rho$ is large, for instance, $r=8$. In such cases, a total of 128 parameters are involved in the optimization Eq.~\eqref{convexhull}, and the numerical evaluation becomes extremely challenging. 

\begin{figure}[t]
    \centering
    \includegraphics[width=0.38\textwidth]{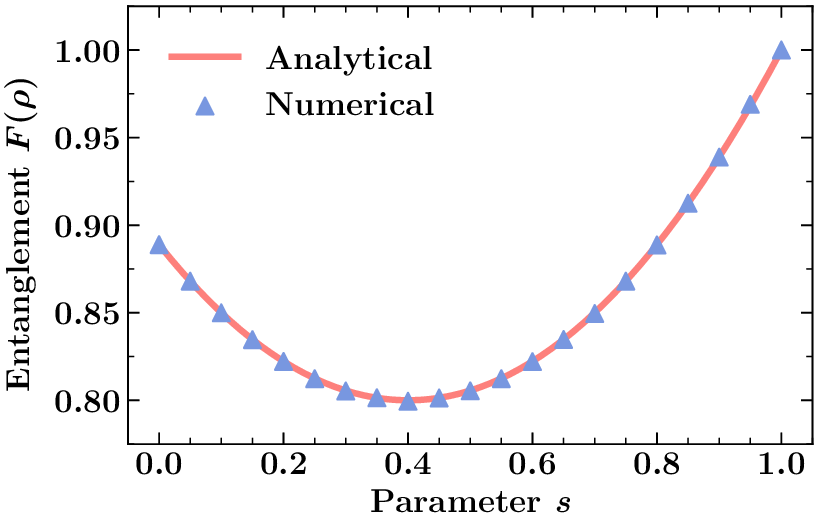}
    \caption{Solid line: Analytical solution of concurrence fill for the states $\rho(s)=s|\text{GHZ}\rangle\langle \text{GHZ}|+(1-s)|W\rangle\langle W|$, with $s$ from 0 to 1. Triangle marks: Numerical results of evaluating Eq.~\eqref{conconjugate} for the same states.}
    \label{fig:task2}
\end{figure}

An alternative approach was given by Eisert, Brand\~ao, and Audenaert \cite{eisert2007}. Recognizing the convex property of Eq.~\eqref{convexhull}, they applied the Legendre transform twice to obtain:
\begin{equation}\label{conconjugate}
    F(\rho)=\sup_{X}\inf_{\psi}\Big\{\text{Tr}\big[X(\rho-|\psi\rangle\langle\psi|)\big]+F(\psi)\Big\},
\end{equation}
where the interior infimum is taken over all possible three-qubit pure states $\psi$ with 14 parameters, while the exterior supremum is taken over all possible $8\times8$ Hermitian matrices $X$. At first glance, Eq.~\eqref{conconjugate} appears more complicated than \eqref{convexhull}, involving both ``$\sup$'' and ``$\inf$'' operations. However, the infinitely many possible decompositions $\{p_i,\psi_i\}$ in \eqref{convexhull} are eliminated. Furthermore, the dimension of the Hermitian $X$ space for the exterior supremum, typically 64 for three-qubit systems, can be significantly reduced by the symmetry of the state $\rho$. This greatly simplifies the numerical task, as demonstrated by Ryu, Lee, and Sim \cite{ryu2012}. As an example of the method, the concurrence fill measure for the mixture  $\rho(s)=s|\text{GHZ}\rangle\langle\text{GHZ}|+(1-s)|W\rangle\langle W|$ is shown in Fig.~\ref{fig:task2}. The definitions of the Greenberger-Horne-Zeilinger (GHZ) state and the $W$ state can be found in \cite{dur2000}. The analytic solution for the mixture is $F(s)=(5s^2-4s+8)/9$, provided by Lohmayer {\it et~al.} \cite{lohmayer2006}. For this state, the dimension of the exterior supremum for $X$ in \eqref{conconjugate} is reduced to 8 (compared to 64) due to the state's symmetries. 

This {\it minmax} approach, on the other hand, has its own limitations. When the symmetry of the mixed state $\rho$ fails to effectively reduce the external dimension of $X$, the method becomes less efficient. 

In this work, to investigate three-qubit ESD, we will utilize a combination of the two approaches. Specifically, we employ different methods for different states, choosing the most suitable approach for each mixed state to achieve optimal numerical performance. It turns out that the problem under discussion can be effectively addressed by combining these two approaches. Moreover, the Nelder-Mead algorithm in Fortran \cite{press1992} is applied to calculate $F(\rho)$ in both approaches.

{\it Spontaneous emission for a three-qubit system}.---In the three-qubit model, we consider three additional cavity-qubits to create independent spontaneous emissions. The system-qubits are labeled by $A$, $B$, and $C$, while the cavity-qubits are labeled by $a$, $b$, and $c$. We allow the six qubits to interact with each other only in pairs, e.g., $A$ with $a$, and the other pairs similarly, as illustrated using $A$ and $a$ as an example:
\begin{equation}\label{evolution}
    \begin{split}
        &|1\rangle_A|0\rangle_a\xrightarrow{\ t/\tau\ }p(t)|1\rangle_A|0\rangle_a+q(t)|0\rangle_A|1\rangle_a,\\ &|0\rangle_A|0\rangle_a\xrightarrow{\ t/\tau\ }|0\rangle_A|0\rangle_a,
    \end{split}
\end{equation}
where the coefficients are $p(t)=(e^{-t/\tau})^{1/2}$ and $q(t)=(1-e^{-t/\tau})^{1/2}$. The interaction is widely recognized as amplitude damping, with a notable example being the zero-temperature spontaneous emission for two-level atoms, where an excited atom in the state $|1\rangle$ eventually emits a photon and evolves into the ground state $|0\rangle$. In our example, the three cavity-qubits always start from the state $|000\rangle_{abc}$. Although the three system-qubits $ABC$ have no direct interactions among themselves, they are initially prepared in a genuinely entangled pure state, to be determined separately below. After an infinite time, the system-qubits evolve into the product state $|000\rangle_{ABC}$, with no entanglement. ESD refers to the scenario when entanglement can vanish earlier, within a finite time.

\begin{figure}[t]
    \centering
    \includegraphics[width=0.35\textwidth]{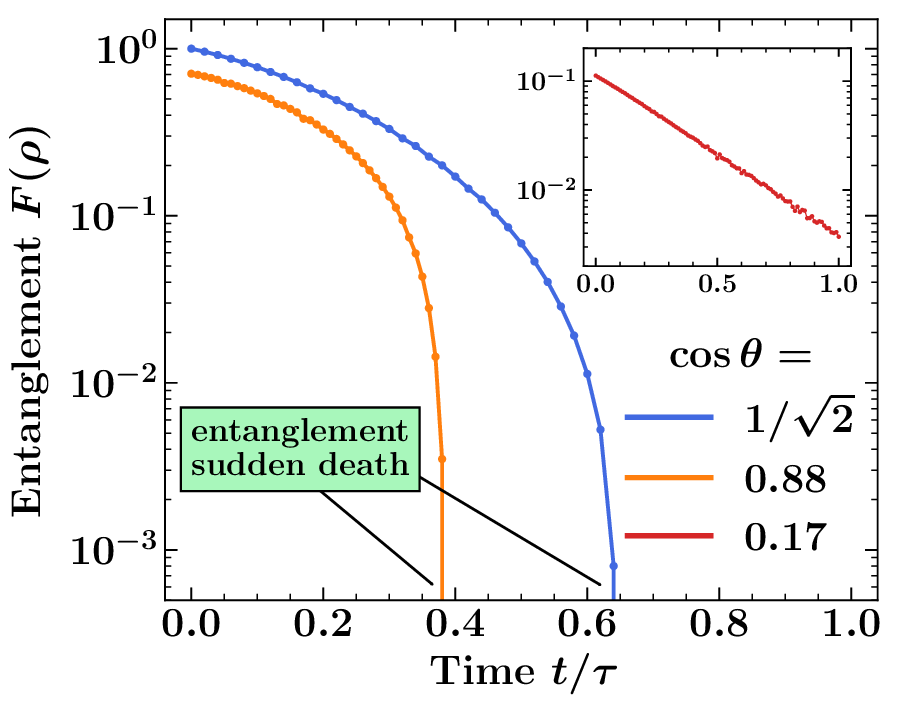}
    \caption{Dynamics of concurrence fill with the initial state Eq.~\eqref{ghz}. The parameters are $\cos\theta=1/\sqrt{2}$ (blue), 0.88 (orange), and 0.17 (red). The results are presented using a logarithmic vertical scale. The blue and orange curves exhibit ESD, while the red curve (inset) shows exponential decay when converted into a linear scale.}
    \label{fig:ghz}
\end{figure}

{\it Different initial states}.---We first prepare the initial state for the system as
\begin{equation}\label{ghz}
    \begin{split}
        |\text{GHZ}(\theta)\rangle=\cos\theta|111\rangle+\sin\theta|000\rangle,
    \end{split}
\end{equation}
a generalized form of the GHZ state, which is then subjected to the amplitude damping channel Eq.~\eqref{evolution}. The rank of the system's density matrix $\rho$ is $r=8$, making the {\it cutoff} approach less efficient. However, $\rho$ exhibits strong symmetries, enabling the exterior dimension of $X$ in \eqref{conconjugate} to be reduced to 6. Therefore, for this state, we choose the {\it minmax} approach to numerically evaluate the genuine entanglement for the system-qubits $ABC$.

The results are provided in Fig.~\ref{fig:ghz}, with three different values of $\cos\theta$: $1/\sqrt{2}$ (blue), 0.88 (orange), and 0.17 (red). These results are displayed using a logarithmic vertical scale. The inset reveals a straight line for the red curve, corresponding to a typical exponential decay when plotted on a linear scale. This suggests no ESD for $\cos\theta=0.17$. In contrast, the blue and orange curves abruptly reach negative infinity before $t/\tau=0.65$ and $t/\tau=0.4$, respectively. Their constant negative infinities in the log scale correspond to constant zero values for those two curves on a linear scale. Consequently, they indicate the occurrence of ESD for the genuine tripartite entanglement in the system.

A side remark can be made: the initial $\text{GHZ}(\theta)$ mixed state for the system-qubits $ABC$ in \eqref{ghz} remains what is called an $X$-form density matrix \cite{yu2007} throughout the amplitude damping interactions with the cavity-qubits $abc$. A separate GME measure, called {\it genuine multipartite concurrence} and labeled GMC \cite{ma2011} was earlier found by Hashemi-Rafsanjani {\it et al.} \cite{hashemi2012} to have a closed form for $X$-matrices, even if they are mixed states. Here, we apply their results to verify the outcome in Fig.~\ref{fig:ghz}.

For the initial state Eq.~\eqref{ghz}, the GMC measure for the entanglement of the system-qubits is
\begin{equation}\label{GMCformula}
    \text{GMC}=\max\big[0,\ p^3(t)|\sin2\theta| -6 p^3(t)q^3(t) \cos^2\theta\ \big],
\end{equation}
with which it can be confirmed that ESD begins at the time specified by:
\begin{equation}
    |\tan\theta|=3\left(1-e^{-t/\tau}\right)^{3/2}.
\end{equation}
Since the value on the right-hand side is in the range $[0,3)$, it implies that, in order for the state \eqref{ghz} to exhibit ESD, one must choose $|\tan\theta|<3$, or equivalently $|\cos\theta|>1/\sqrt{10}\approx0.316$.

\begin{figure*}[t]
    \centering
    \includegraphics[width=0.7\textwidth]{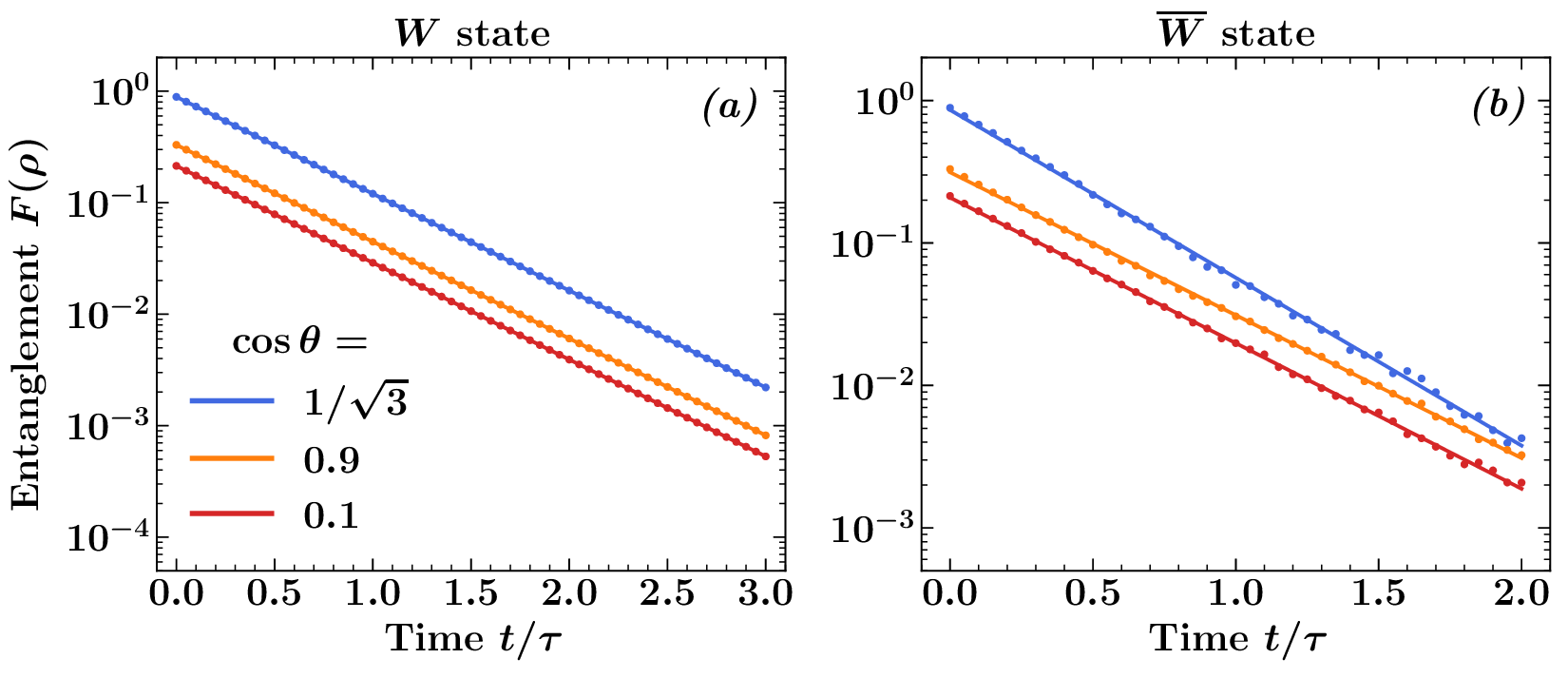}
    \caption{Dynamics of concurrence fill with the initial state Eq.~\eqref{w}. The parameters are $\cos\theta=1/\sqrt{3}$ (blue), 0.9 (orange), and 0.1 (red). The results are presents using a logarithmic vertical scale. No ESD is exhibited.}
    \label{fig:wwbar}
\end{figure*}

Both concurrence fill and GMC are GME measures. Therefore, when one exhibits ESD, so does the other one. Based on this information, we find that our results in Fig.~\ref{fig:ghz} corroborate the ESD predictions made using GMC. Specifically, when $\cos\theta=1/\sqrt{2}$ and $0.88$, which are both greater than $1/\sqrt{10}$, we observe ESD. When $\cos\theta=0.17$, ESD does not occur.

Moving forward, we examine the following generalized $W$-type initial states for the system:
\begin{equation}\label{w}
    \begin{split}
        &|W(\theta)\rangle=\cos\theta|100\rangle+\dfrac{1}{\sqrt{2}}\sin\theta(|010\rangle+|001\rangle),\\
        &|\overline W(\theta)\rangle=\cos\theta|011\rangle+\dfrac{1}{\sqrt{2}}\sin\theta(|101\rangle+|110\rangle).
    \end{split}
\end{equation}
The $W(\theta)$ and $\overline W(\theta)$ states correspond to single-excitation and double-excitation cases of the genuinely entangled $W$ class \cite{dur2000}. They are connected by local spin-flip operations but exhibit different dynamics when subjected to the amplitude damping channel \eqref{evolution}. The parameter $\theta$ introduces bias on the first qubit. 

It is important to emphasize that the density matrices for the initial $W(\theta)$ and $\overline W(\theta)$ states are no longer in the form of $X$-matrices, rendering the GMC formula from \cite{hashemi2012} inapplicable for examining the existence of ESD. Therefore, numerical techniques are necessary. The symmetries of the density matrices $\rho$ are not significant enough in the sense that the exterior dimension of $X$ in Eq.~\eqref{conconjugate} can only be reduced to 12, making the {\it minmax} approach less efficient and yielding inaccurate results. Conversely, the density matrices have a relatively small rank, $r=5$. For this reason, we opt for the {\it cutoff} approach to evaluate the entanglement for the initial states \eqref{w}. 

The results are presented in Fig.~\ref{fig:wwbar} with $\cos\theta=1/\sqrt{3}$ (blue), $0.9$ (orange), and $0.1$ (red), using a logarithmic vertical scale for both states. Surprisingly, the curves for both initial states remain straight lines, regardless of the chosen value of $\theta$, indicating exponential decays on a linear scale. Therefore, no ESD can be observed for these initial states.

A question may arise: Fig.~\ref{fig:wwbar} only displays the absence of ESD at times earlier than $t/\tau=3$ and $t/\tau=2$. How can one be certain that ESD cannot occur at much later times, such as $t/\tau=10$ or even $t/\tau=100$? We emphasize that the intrinsic time scale of our interaction is $\tau$. Consequently, when varying the $\cos\theta$ value for these states, it is reasonable to expect that ESD, if present, would manifest as an abrupt logarithmic drop at some time not far from $t/\tau\sim 1$.  Clearly, no such behavior is observed for the three $\cos\theta$ values selected. This point can be further clarified when we discuss the origin of ESD in the following section.

{\it A state that triggers ESD}.---A hidden competing process should be pointed out for the state in Eq.~\eqref{ghz}. When the $|111\rangle$ component has a higher amplitude, ESD occurs earlier, as observed in the orange curve in Fig.~\ref{fig:ghz} compared to the blue curve. However, if the amplitude of $|111\rangle$ is below a threshold ($1/\sqrt{10}$ in this case), we can say that ESD occurs later than infinite time, implying that ESD never happens, as seen in the red curve. Such competition is absent for the initial states in Eq.~\eqref{w}, regardless of the chosen parameter $\theta$. This highlights the $|111\rangle$ state as the unique state that triggers ESD occurrence in the current model.

\begin{figure}[b]
    \centering
    \includegraphics[width=0.35\textwidth]{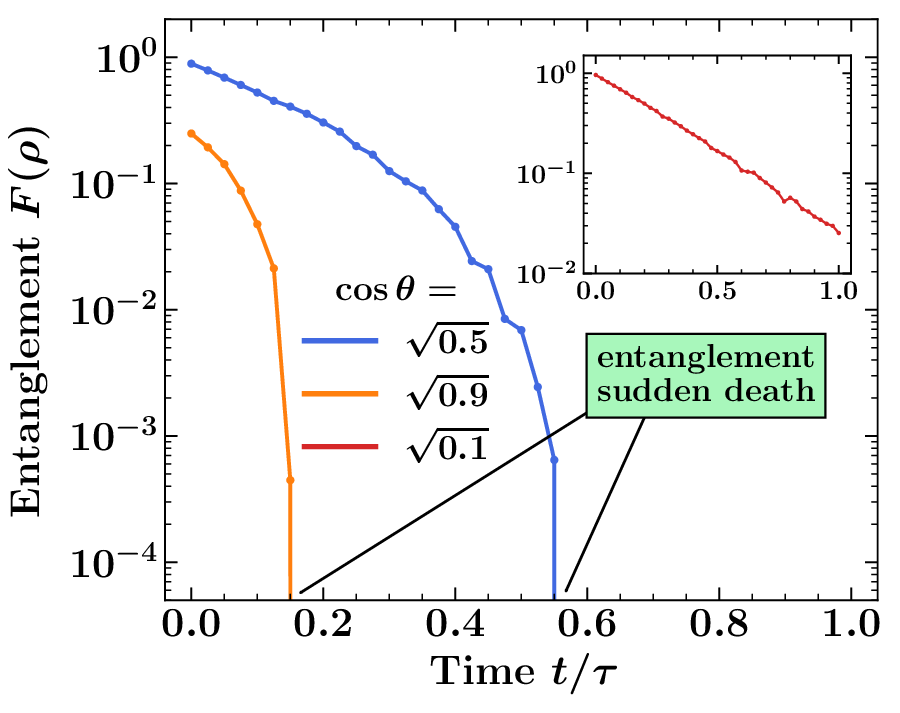}
    \caption{Dynamics of concurrence fill with the initial state Eq.~\eqref{sigma}. The parameters are $\cos\theta=\sqrt{0.5}$ (blue), $\sqrt{0.9}$ (orange), and $\sqrt{0.1}$ (red). The results are presented using a logarithmic vertical scale. The blue and orange curves exhibit ESD, while the red curve (inset) shows exponential decay when converted into a linear scale.}
    \label{fig:sigma}
\end{figure}

To further emphasize the critical role of $|111\rangle$, we explore an additional step. We speculate that even when $|111\rangle$ is superposed with any other state, a similar competing process of ESD will remain. To verify this conjecture, we superpose $|111\rangle$ with the all-symmetric state $|W\rangle=(|100\rangle+|010\rangle+|001\rangle)/\sqrt{3}$, to obtain
\begin{equation}\label{sigma}
    \begin{split}
        |\Sigma(\theta)\rangle=\cos\theta|111\rangle+\sin\theta|W\rangle.
    \end{split}
\end{equation}
Here, the rank of the density matrices is $r=8$. For the same reason as in the generalized GHZ state \eqref{ghz}, we employ the {\it minmax} approach for this state. Due to the density matrix's symmetries, the exterior dimension of $X$ in \eqref{conconjugate} is reduced to 10.

The results are presented in Fig.~\ref{fig:sigma}. The competing process is quite obvious, even though the exact threshold value for $\cos\theta$ remains unknown in this case. The orange curve with $\cos\theta=\sqrt{0.9}$ exhibits ESD starting relatively early, while the ESD for the blue curve with $\cos\theta=\sqrt{0.5}$ appears later. For the red curve with $\cos\theta=\sqrt{0.1}$, no ESD is observed, as indicated by the straight line in the log scale. These results further support the notion that the presence of the $|111\rangle$ state in the initial condition enables ESD to occur.

What happens when we take the state $|111\rangle$ alone as the initial state? The answer is rather straightforward. As $|111\rangle$ is an unentangled product state, the entanglement dynamics will remain constantly zero. Alternatively, one argues that ESD is so potent that it occurs immediately, right from the start at $t/\tau=0$.

{\it Discussions of the results}.---Our work focuses on three-qubit systems, but the same conclusion can be observed in two-qubit ESD examples, where the entanglement can be easily calculated using the closed-form concurrence formula \cite{hill1997}. In this case, the ``trigger state'' in the initial condition is $|11\rangle$. These discoveries lead to the natural speculation that our result is universal---ESD can only be triggered by the initial dominance of the all-excited state $|11\cdots1\rangle$ in a system with an arbitrary number of qubits. 

We can now revisit the assumption posed at the beginning of this work. It turns out that ESD of genuine entanglement relies heavily on the initial amplitude of the unique ``ESD-trigger state'' $|11\cdots1\rangle$. However, as the number of qubits in the system increases, the dimension of the Hilbert space grows much faster, causing the trigger state to occupy a progressively smaller portion of the entire space. Consequently, the likelihood of observing ESD diminishes rapidly. This observation contradicts the intuition that GME is at greater risk of experiencing sudden death.

Finally, we emphasize here that our discoveries only focus on the amplitude damping channel Eq.~\eqref{evolution}, which represents the common process of spontaneous emission occurring ubiquitously. However, given the numerical techniques employed in this work, it would be intriguing to explore cases involving other channels, such as depolarizing and dephasing channels, for which two-qubit ESD has previously been observed. An compelling geometric illustration of the effects of different environments on two-qubit ESD was studied in \cite{cunha2007}. Furthermore, investigating ESD in qudit systems could also be an interesting area of research.

We thank Professors W.A.~Coish, L.~Davidovich, D.F.V.~James, X.-F.~Qian, and T.~Yu for valuable discussions. Financial support was provided by National Science Foundation Grants No.~PHY-1505189 and No.~PHY-1539859 (INSPIRE), and a competitive grant from the University of Rochester. Calculations were performed on the BlueHive supercomputing cluster at the University of Rochester.

\end{document}